\documentclass[twocolumn,showpacs,preprintnumbers,amsmath,amssymb,amsfots]{revtex4}

\usepackage{graphicx}
\usepackage{dcolumn}
\usepackage{bm}
\usepackage{mathrsfs}

\newcommand{\lag}{\ensuremath{\mathscr{L}}}        
\newcommand{\ld}{\ensuremath{{\mathcal{L}}_{\bm{\xi}}}} 
\def\dv{\sqrt{| g |}\ d^n x}
\def\nn{\nonumber\\}
\def\st{space-time}

\def\F#1#2{\ensuremath{F_{#1#2}}}           
\def\cd#1{\ensuremath{\nabla_{#1}}}          
\def\pd#1{\ensuremath{\partial_{#1}}}        
\def\tm#1#2{\ensuremath{T_{\mathscr{M}}^{#1#2}}} 
\def\tc#1#2{\ensuremath{T_{\mathscr{C}}^{#1#2}}}
\def\trad#1#2{\ensuremath{T_{c}^{#1#2}}}
\def\t#1#2{\ensuremath{{T^{#1#2}}}}
\def\jota#1{\ensuremath{{\mathcal{J}_{\bm \xi}^{#1}}}}
\def\eq#1{(\ref{#1})}

\begin{document}


\title{The electromagnetic  energy-momentum tensor}

\author{Ricardo E.\ Gamboa~Sarav\'{\i}}

 \email{quique@venus.fisica.unlp.edu.ar}
\affiliation{Departamento de F\'\i sica, Universidad Nacional de
La Plata\\ C.C. 67, 1900 La Plata, Argentina
}%

\date{\today}

\begin{abstract}
We clarify the relation between   canonical  and  metric
energy-momentum
 tensors. In particular, we show that  a natural definition arises from
 Noether's Theorem which
 directly leads to a symmetric and  gauge invariant
 tensor for
electromagnetic field theories on an arbitrary \st\ of any
dimension.

\end{abstract}

\maketitle

For many decades a suitable definition for the  energy-momentum
tensor has been under investigation. This is not 
a technical point, not only because $T^{ab}$ should  provide
meaningful physical conserved quantities, but also because it is
the source of Einstein's gravitational field equations.

  In flat \st,  the canonical energy-momentum tensor arises
from Noether's Theorem by considering
 the conserved currents associated to  translation invariance.
However,  only for scalar fields the energy-momentum tensor
constructed in this way turns out to be symmetric. Moreover, for
Maxwell's theory, it breaks the gauge symmetry. Of course, it is
possible to correct it by a symmetrization procedure \cite{bel},
although this looks as an {\em ad hoc}  prescription.

On the other hand, a completely different approach leads to the
metric energy-momentum tensor (see for example \cite{haw}) which
is, by definition, symmetric and gauge invariant.

The aim of this paper is to clarify the relation between these
tensors.

Let us  consider a field theory where the Lagrangian \lag\ is a
local function of $\F a b$, the exterior derivative $\partial_a
A_b-\partial_b A_a$, of a one-form field  $A_b$, and  the metric
tensor $g_{ab}$, defined on a (semi)Riemannian  manifold  of
dimension $n\geq2$.

The field equations  are obtained by requiring that the action
\begin{equation}\label{action}
S=\int_\Omega \lag(\F a b,g_{ab})\ \dv,
\end{equation}
 be stationary under arbitrary variations of the fields $\delta
A_b$ in the
interior of any compact region $\Omega$. Thus, one obtains
\begin{equation}\label{eqm}
\cd a \left(\frac{\partial\lag}{\partial \F ab} \right)=0,
\end{equation}
where \cd a is the covariant derivative associated to the
Levi-Civita connection.

Needless to say, that even in flat \st\ we are allowed to use
curvilinear coordinates, so the action (\ref{action}) must be
invariant under general coordinate transformations. This requires
$\lag$ to be a scalar function. Thus, its Lie derivative  with
respect to any vector field $\xi^a$, $\ld \lag$, must satisfy
\begin{equation}\label{lie}
\ld \lag -\cd a(\lag)\ \xi^a=0.
\end{equation}

Now, taking into account that the Lagrangian $\lag$ depends on
the coordinates only through the tensor fields $ \F ab$ and $g_{ab}$,
we have
\begin{eqnarray}\label{4}
\ld \lag= \frac{\partial\lag}{\partial \F ab}\ \ld  \F ab
+\frac{\partial\lag}{\partial g_{ab}}\ \ld
g_{ab}.
\end{eqnarray}
But, for any   tensor field \F ab of the type $(0,2)$, it holds
\begin{eqnarray}\label{222}
\ld \F ab = \xi^c\ \cd c \F ab +  \F cb \cd a \xi^c+ \F ac \cd b
\xi^c\ ;
\end{eqnarray}
thus
\begin{eqnarray}\label{333}
 \frac{\partial\lag}{\partial \F ab}\ \ld  \F ab
= 2\ \frac{\partial\lag}{\partial \F ab}
\left( \cd a \F cb\ \xi^c+\F cb\ \cd a \xi^c\right)\nn
= 2\ \frac{\partial\lag}{\partial \F ab} \cd a \left( \F cb\ \xi^c\right),
\end{eqnarray}
where we have used the identity
\begin{eqnarray}\label{bianchi}
 \cd a \F bc + \cd b \F ca + \cd c \F ab=0\ ,
\end{eqnarray}
($d\bm F= d(d\bm A)=0$), and the obvious antisymmetry of  the
$(2,0)$ tensor field $\tfrac{\partial\lag}{\partial \F ab}$
.

Now, for fields satisfying the equations of motion (\ref{eqm}),
\eq{333} reads
\begin{eqnarray}\label{7}
 \frac{\partial\lag}{\partial \F ab}\ \ld \F ab
= 2\ \cd a \left(\frac{\partial\lag}{\partial \F ab}  \F cb\
\xi^c\right).
\end{eqnarray}

So,  from (\ref{lie}), (\ref{4}) and (\ref{7}), we get for any
vector field $\xi^a$
\begin{eqnarray}\label{8}
 \nabla_a\left(2\ \frac{\partial\lag}{\partial \F ac}\ F^b_{\ c}\ \xi_b
  \right) + \frac{\partial\lag}{\partial g_{ab}}\ \ld g_{ab}
-\cd a(\lag)\ \xi^a=0.
\end{eqnarray}
 Moreover, applying \eq{222} to the metric tensor, we have
\begin{equation}
\ld g_{ab}= \nabla_a \xi_b+\nabla_b \xi_a ,
\end{equation}
and so, \eq{8} can be rewritten as
\begin{eqnarray}\label{88}
 \nabla_a\left(2\ \frac{\partial\lag}{\partial \F ac}\ F^b_{\ c}\ \xi_b
 -g^{ab} \lag\  \xi_b \right)\nn +\left( \frac{\partial\lag}{\partial g_{ab}}+\frac{1}{2}\
 g^{ab} \lag  \right) \ld g_{ab}=0.
\end{eqnarray}
We define the ``true" canonical   energy-momentum tensor as
\begin{eqnarray}
\tc ab:= -2\ \frac{\partial\lag}{\partial \F ac}\ F^b_{\ c}+
g^{ab} \lag , \label{tc}
\end{eqnarray}
and the metric   one as
\begin{eqnarray}
\tm ab:= 2\ \frac{\partial\lag}{\partial g_{ab}}+ g^{ab} \lag
.\label{tm}
\end{eqnarray}
Notice that, by definition, $\tm ab$ is a symmetric $(2,0)$
tensor.

In terms of these tensors (\ref{88}) reads
\begin{eqnarray}\label{magic}
 \cd a \left(\tc ab \xi_b \right) -\frac{1}{2}\ \tm ab\ \ld g_{ab}=0.
\end{eqnarray}
This last equation, a rewritten form of \eq{lie}, which holds for any vector field $\xi^a$,
has several important consequences. In fact,
we shall obtain all the results of this work by using it in four different ways:

i) Let us  restrict attention to the case where  $\xi^a$ is a
Killing vector field, i.e. a generator of an infinitesimal
isometry, so $\ld g_{ab}=\cd a \xi_b+\cd b \xi_a=0$. From
(\ref{magic}), one directly   obtains the Noether current  $\jota
a $ associated to this symmetry
\begin{eqnarray}\label{Noether}
\cd a \jota a = \cd a (\tc ab  \xi_b)=0,
\end{eqnarray}
for, in this case, the last term in \eq{magic}
clearly vanishes. But this is just the beginning.

 ii) At any point of the manifold we can choose  Riemannian  normal coordinates $x^\alpha$
 (i.e., a local inertial coordinate system). Moreover, we can choose for $\xi_b$ any set of $n$
linear independent covectors with constant components in this coordinate system. For instance,
the dual basis covectors $dx^{\alpha}_b$. So,  \eq{magic}
reads
\begin{eqnarray}
 \pd \alpha (\tc \alpha\beta )\ \xi_\beta
 +  \tc \alpha\beta \pd \alpha \xi_\beta
 - \tm \alpha\beta\ \pd \alpha \xi_\beta=
 \pd \alpha (\tc \alpha\beta )\ \xi_\beta =0 \nn
\end{eqnarray}
because of  the vanishing of Christoffel symbols and  partial
derivatives of $\xi_\beta$. Hence, we get $\cd \alpha \tc
\alpha\beta = \pd \alpha \tc \alpha\beta =0$. But this is a tensor
relation, then \footnote{ Of course, it also follows directly from
the equations of motion (\ref{eqm}), for
$
\cd a \tc ab= 2 \frac{\partial\lag}{\partial\F ac} \cd a F^b_{\ c}
- g^{ab} \cd a \lag
=2 \frac{\partial\lag}{\partial\F ac} \cd a F^b_{\ c}
- g^{ab} \frac{\partial\lag}{\partial\F cd} \cd a \F cd
=2 \frac{\partial\lag}{\partial\F ac} \cd a F^b_{\ c}
-2 \frac{\partial\lag}{\partial\F ac} \cd a F^b_{\ c}=0
$
, where we have used the identity (\ref{bianchi}).}
\begin{eqnarray}
\cd a \tc ab =0.
\end{eqnarray}

iii) Coming back to (\ref{magic}), we  rewrite it as
\begin{eqnarray}\label{12}
\nabla_a\left(\ (\tc ab -\tm ab)\ \xi_b \right)
+ \nabla_a(\tm ab)\ \xi_b=0.
\end{eqnarray}

Now, we integrate (\ref{12}) over any compact region $\Omega$,
taking arbitrary vector fields $\xi^a$ vanishing everywhere except
in  its  interior. The first contribution may be transformed into
an integral over the boundary which vanishes, as $\xi^a$ is zero
there. Since the second term must therefore be zero for arbitrary
$\xi^a$, it follows that
\begin{eqnarray}\label{127}
\cd a \tm ab=0.
\end{eqnarray}

iv) Now, coming back to (\ref{magic}) written as in \eq{12},  we
see that the diffeomorphism invariance of the action yields not
only  $\cd a \tc ab=\cd a \tm ab =0$, but also
\begin{eqnarray}\label{22}
\cd a\left(\ (\tc ab -\tm ab)\ \xi_b \right)=(\tc ab -\tm ab)\cd a \xi_b=0,
\end{eqnarray}
for any covector field $\xi_b$. Therefore, since $\cd a \xi_b$ is arbitrary,
we conclude that both tensors  coincide
\begin{eqnarray}
\tc ab =\tm ab.
\end{eqnarray}

We have thus shown that
\begin{eqnarray}
\cd a \tc ab=0,\ \ \cd a \tm ab=0\ \text{and}\ \ \tc ab =\tm ab,
\end{eqnarray}
follow as a consequence of the diffeomophism invariance of the
action.

Some comments are in order. We want to point out that \tc ab\ has
nothing to do with Killing vectors. \tc ab depends only on the
fields, their derivatives and the metric, and $\cd a \tc ab =0 $
is always true, even when the metric has no isometry at all. But,
of course,
 a tensor by itself does not give rise to any
conserved quantity \footnote{ For $\cd a \t ab
=\dfrac{\partial_a(\sqrt{-g}T^{ab})}{\sqrt{-g}}+ T^{ac}\
\Gamma^b_{\ ca}$} so, in order to construct conserved quantities,
it is necessary to have a Killing vector at hand to construct the
current $\jota a = \tc ab \xi_b$.

The \tc ab we define in (\ref{tc}) 
arises  naturally from Noether's theorem taking into account that
the Lagrangian  is a local function of \F ab. It is important to
realize that, as shown by \eq{Noether},   if \st\ admits a Killing
vector we obtain from \tc ab a conserved current \jota a. Thus,
for instance,  the $n(n+1)/2$ currents in Minkowski \st\ are
obtained from \tc ab, by contracting it with the corresponding
Killing vectors.

In Minkowski space-time,
the canonical energy-momentum tensor is defined as (see for example
\cite{land,jack})
\begin{eqnarray}\label{trad}
\trad ab:= -\frac{\partial\lag}{\partial\ \!\cd a A_c}\ \nabla^b
A_c +  g^{ab} \lag,
\end{eqnarray}
perhaps as a simple generalization of its expression for scalar
fields.
 It seems to
us that this definition
 is in some sense unnatural, for
\lag\ depends on the $n(n-1)/2$ components of $d\bm A$ and not on
the $n^2$ derivatives $\cd a A_b$.

\trad ab\  defined as in \eq{trad} is neither symmetric nor gauge
invariant. Of course, in flat space-time,  $\cd a \trad ab=0$
holds. But, it is worthwhile noticing that this is not even true
for curved \st, for
\begin{eqnarray}\label{R}
\cd a \trad ab= -2 \frac{\partial\lag}{\partial\F ac} \cd a
\nabla^b A_c + g^{ab} \cd a \lag\nn =-2
\frac{\partial\lag}{\partial\F ac} \cd a \nabla^b A_c + g^{ab}
\frac{\partial\lag}{\partial\F cd} \cd a \F cd\nn =-2\
\frac{\partial\lag}{\partial\F ac} \cd a \cd c A^b
 =-\frac{\partial\lag}{\partial\F ac}\  R^b_{\ dac}\ A^d,
\end{eqnarray}
where, again, we have used the identity (\ref{bianchi}), and
$R^b_{\ dac}$ is the Riemann curvature tensor. So, not only \trad
ab\ is neither symmetric nor gauge invariant, but also $\cd a
\trad ab$ vanishes only when space-time is flat.

Moreover, in flat \st, for a Killing  field $\xi_b$\ it holds $\cd
a (\trad ab \xi_b)= \trad{[a}{b]} \cd a \xi_b$, so the current
$\trad ab \xi_b$ is conserved only for constant $\xi_b$, for
  \trad ab is not symmetric. Then
we get  from \trad ab only  $n$ currents associated with the
constant Killing vectors (translations). A similar result holds
for
 curved \st , even though $\cd a \trad ab \neq 0$.
In fact, if there exists a constant Killing vector ($\cd a
\xi^b=0$) we have
\begin{eqnarray}
\cd a (\trad ab \xi_b)= \cd a \trad ab\ \xi_b
=\frac{\partial\lag}{\partial\F ac}\ A^d  R^b_{\ dac}\ \xi_b \nn
=2 \frac{\partial\lag}{\partial\F ac}\ A^d\  \cd c \cd a \
\xi_d=0\ ,
\end{eqnarray}
and so, we get a conserved current for each constant Killing
vector $\xi_b$.

Notice that $\tc ab =\tm ab$ means that for any scalar Lagrangian
depending on the tensor fields $ \F ab$ and $g_{ab}$
\begin{eqnarray}\label{off}
\frac{\partial\lag}{\partial \F ac}\ F^b_{\ c}
=-\frac{\partial\lag}{\partial g_{ab}}\ .
\end{eqnarray}
 Moreover, as the right hand-side
is a symmetric tensor field, so  is the left hand-side. It is
worth bearing in mind that \eq{off} holds off-shell too. In fact,
 from \eq{lie} we have for any field configuration
\begin{eqnarray}
\ld \lag -\cd b(\lag)\ \xi^b=\frac{\partial\lag}{\partial \F ac}\
(\ld  \F ac-\cd b \F ac\ \xi^b)\nn +\frac{\partial\lag}{\partial
g_{ab}}\ \ld g_{ab}
 =2\left(\frac{\partial\lag}{\partial \F ac}\
F^b_{\ c} +\frac{\partial\lag}{\partial g_{ab}}\right)\cd a
\xi_b=0\ ,
\end{eqnarray}
and the vector field $\xi^b$ is completely arbitrary.

 For the sake of clarity, let us consider
 the translation invariance in flat \st . Taking cartesian
coordinates we compute $\pd a \lag$ in two different ways. First,
thinking of \lag\ as function of \F ab, as it actually  is, we get
\begin{eqnarray}
\pd a \lag = \frac{\partial\lag}{\partial\F cd}\ \pd a \F cd\nn
=2\ \frac{\partial\lag}{\partial\F cd}\ \pd c \F ad =  2\ \pd c
\left(\frac{\partial\lag}{\partial\F cd}\  \F ad \right).
\end{eqnarray}
 In the second step we used $\pd
{[a}\F a{b]}=0$ and the field equations in the third. Thus we get
the conservation law $\pd a \tc ab=0$.

On the other hand, if we consider \lag \ a function of $\pd a A_b$
\begin{eqnarray} \pd a \lag=
\frac{\partial\lag}{\partial\ \! \pd c A_d}\ \pd a \pd c A_d\nn
=\frac{\partial\lag}{\partial\ \! \pd c A_d}\ \pd c \pd a A_d =\pd
c\left(\frac{\partial\lag}{\partial\ \! \pd c A_d}\  \pd a
A_d\right),
\end{eqnarray}
where we have commuted the partial derivatives. And this is the
conservation $\pd a \trad ab =0$.

Clearly, the first computation still holds in curved \st. But the
second one fails, for covariant derivatives acting on one-forms do
not commute, and we get \eq{R}.

For scalars fields,  our arguments remain valid. The only change
to be done is the definition   $\tc ab:=-
\dfrac{\partial\lag}{\partial\ \pd a \phi}\
\partial^b \phi +g^{ab} \lag$.  Equation \eq{magic} still holds and all the results follow
as above. For general tensor fields, the dependence of the
Lagrangian on the affine connection as well as the
noncommutativity between \cd a\  and \ld\ make the computation
more involved \cite{q}.

Summarizing, we have shown that, properly defined as in \eq{tc},
the canonical energy-momentum tensor \tc ab is symmetric, gauge
invariant and coincides with \tm ab. Moreover, it is the one that
arises  naturally from Noether's  theorem when the metric has
isometries, and all the currents are written as $\jota a = \tc ab
\xi_b$. For these reasons, we call \tc ab the ``true" canonical
energy-momentum tensor.


\begin{acknowledgements}
The author wishes to thank  Jorge Solomin for valuable
discussions. This work was supported in part by CONICET,
Argentina.
\end{acknowledgements}

\end{document}